\documentclass[reprint,amsmath, amssymb, superscriptaddress]{revtex4-1}

\usepackage{graphicx}
\usepackage{dcolumn}
\usepackage{bm}
\usepackage[mathlines]{lineno}
\usepackage[utf8]{inputenc}
\usepackage[T1]{fontenc}
\usepackage{mathptmx}
\usepackage[normalem]{ulem} 	
\usepackage[usenames,dvipsnames]{color}
\usepackage{xcolor} 
\usepackage{SIunits} 

\newcommand{\csamp}[0]{\ensuremath{C_\mathrm{s}}}
\newcommand{\ctot}[0]{\ensuremath{C_\mathrm{tot}}}
\newcommand{\cf}[0]{\ensuremath{C_\mathrm{f}}}
\newcommand{\vs}[0]{\ensuremath{V_\mathrm{s}}}
\newcommand{\vp}[0]{\ensuremath{V_\mathrm{p}}}
\newcommand{\vg}[0]{\ensuremath{V_\mathrm{g}}}
\newcommand{\ctrans}[0]{\ensuremath{C_\mathrm{q}}}
\newcommand{\gams}[0]{\ensuremath{\gamma_\mathrm{s}}}
\newcommand{\gamdc}[0]{\ensuremath{\gamma_\mathrm{g}}}
\newcommand{\gamp}[0]{\ensuremath{\gamma_\mathrm{p}}}

\begin{document}

\title{Quantum Sensors for Microscopic Tunneling Systems}

\author{Alexander Bilmes*}
\affiliation{Physikalisches Institut, Karlsruhe Institute of Technology, 76131 Karlsruhe, Germany}
\author{Serhii Volosheniuk}
\affiliation{Physikalisches Institut, Karlsruhe Institute of Technology, 76131 Karlsruhe, Germany}
\author{Jan David Brehm}
\affiliation{Physikalisches Institut, Karlsruhe Institute of Technology, 76131 Karlsruhe, Germany}
\author{Alexey V. Ustinov}
\affiliation{Physikalisches Institut, Karlsruhe Institute of Technology, 76131 Karlsruhe, Germany}
\affiliation{National University of Science and Technology MISIS, Moscow 119049, Russia}
\affiliation{Russian Quantum Center, Skolkovo, Moscow 143025, Russia}
\author{J\"urgen Lisenfeld}
\affiliation{Physikalisches Institut, Karlsruhe Institute of Technology, 76131 Karlsruhe, Germany}

\date{\today}

\begin{abstract}
	\centering\begin{minipage}{\linewidth}
		\textbf{
The anomalous low-temperature properties of glasses arise from intrinsic excitable entities, so-called tunneling Two-Level-Systems (TLS), whose microscopic nature has been baffling solid-state physicists for decades. TLS have become particularly important for micro-fabricated quantum devices such as superconducting qubits, where they are a major source of decoherence. Here, we present a method to characterize individual TLS in virtually arbitrary materials deposited as thin-films.
The material is used as the dielectric in a capacitor that shunts the Josephson junction of a superconducting qubit. In such a hybrid quantum system the qubit serves as an interface to detect and control individual TLS. We demonstrate spectroscopic measurements of TLS resonances, evaluate their coupling to applied strain and DC-electric fields, and find evidence of strong interaction between coherent TLS in the sample material. Our approach opens avenues for quantum material spectroscopy to investigate the structure of tunneling defects and to develop low-loss dielectrics that are urgently required for the advancement of superconducting quantum computers.
	}
	\end{minipage}
\end{abstract}

\maketitle 
\setlength{\parskip}{-0.25cm}

\subsection*{Introduction} \noindent
We are still lacking an explanation for the behaviour of amorphous materials at low temperatures $<$10 K~\cite{ZellerPohl71,Leggett13}. Why is it that even widely different materials ranging from biatomic glasses to polymers show quantitatively identical properties such as specific heat and thermal conductivity~\cite{FreemanAnderson86}? The Standard Tunneling Model (STM)~\cite{Anderson:PhilMag:1972,Phillips:JLTP:1972} has been a first attempt to explain these universal anomalies on the basis of two-level systems (TLS) believed to arise from the tunneling of atoms between two energetically similar configurations in the disordered lattice structure. While the STM neglects mutual TLS interactions and fails in the intermediate temperature range of 1 to 10 K, refined models include TLS-TLS interactions~\cite{Burin96,LubchenkoWolynes07,CarruzzoYu20}, assume different types of TLS~\cite{SchechterStamp13}, or consider specific dependencies of TLS potential energies~\cite{Karpov83,Buchenau91}. Since insights from experiments on bulk materials were limited to observing the averaged response from large and inhomogeneous ensembles of TLS, their individual properties remained out of reach.\\

This situation has changed with the advent of superconducting qubits that realize well-controllable macroscopic quantum systems with custom-tailored energy spectra and couplings to the environment. Qubits are implemented from electric resonant circuits employing Josephson tunnel junctions that serve as nonlinear inductances to obtain anharmonic potential wells where discrete eigenstates can be selectively addressed. Driven by the desire to realize solid-state quantum information processors, intensive effort went into the development of advanced circuit designs~\cite{wallraff2004, KochTransmon, Wang2015, Lin2018, Earnest2018} and fabrication techniques~\cite{RigettiFab,Houck20} which resulted in a dramatic improvement of device coherence. The entry of commercial enterprises has further accelerated progress, culminating in the demonstration of machine learning algorithms~\cite{Rigetti19}, access to prototype quantum processors via the cloud~\cite{IBM19}, and the achievement of quantum supremacy by controlling a 53-qubit system that could not anymore be simulated efficiently by classical supercomputers~\cite{arute2019quantum}.\\

\begin{figure*}[htbp]
	\begin{center}
	\includegraphics[width=\textwidth]{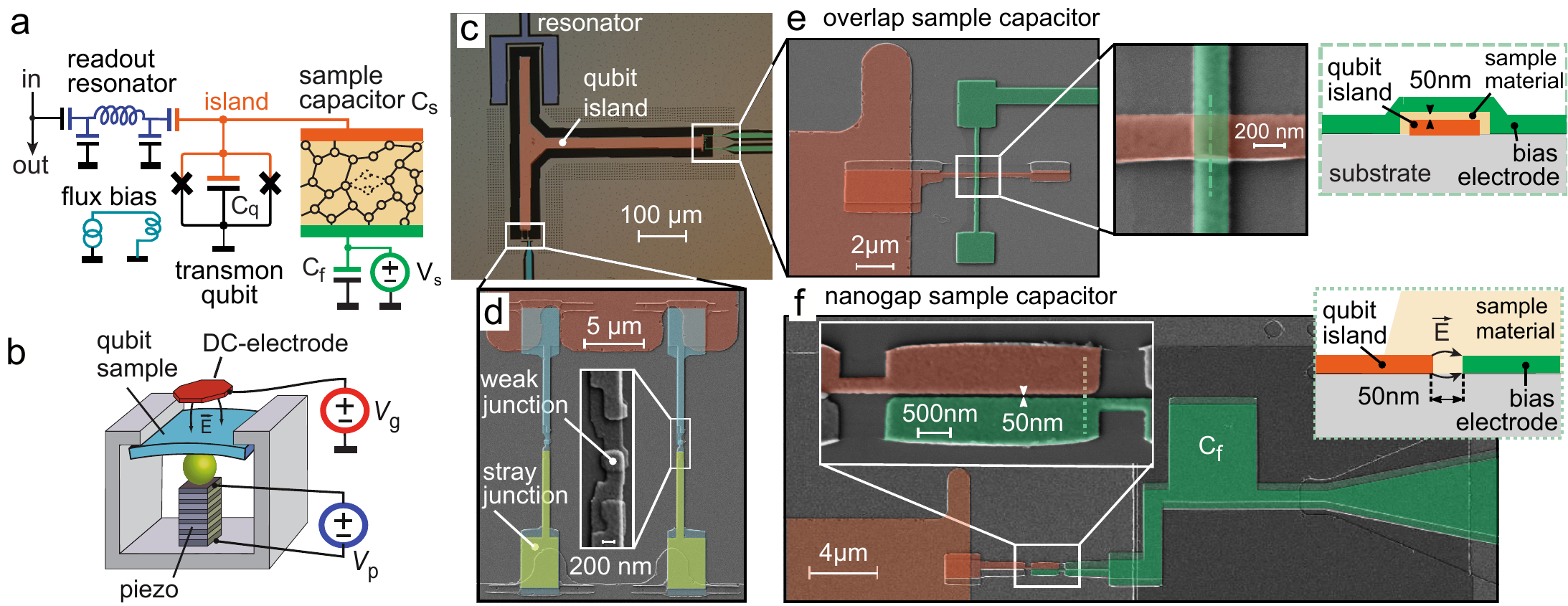}	
	\end{center}
	\caption{
	\textbf{Experimental setup and qubit sample.}
	\textbf{a} Schematic of the transmon qubit circuit to study TLS in deposited materials. The qubit island (red) is connected to ground by an additional small capacitor containing the material to be studied.
	\textbf{b} Setup for tuning TLS by applied mechanical strain and a global DC-electric field.
	\textbf{c} Photograph of the qubit island.
	\textbf{d} DC-SQUID connecting the qubit island to ground, and a zoom onto one of the small Josephson junctions. The large-area stray junctions are highlighted in light green.
	\textbf{e} Sample capacitor in overlap geometry as used in this work. It employs a $50\,\nano\meter$-thick layer of AlO$_x$ as the sample dielectric.
	\textbf{f} An alternative sample capacitor design consists of two coplanar electrodes separated by a so-called 'nanogap' of a few tens of nm. Here, the sample material can be deposited in a last fabrication step (see inset) or left uncovered to study individual TLS in native surface oxides.}
	\label{fig:fig1}
\end{figure*}

Despite these achievements, progress towards truly large-scale quantum processors is still hindered by decoherence, of which the major part is due to losses in dielectric circuit materials~\cite{Muller:2017}.
TLS residing in the tunnel barriers of Josephson junctions and in the native surface oxides of superconducting electrodes may couple via their electric dipole moments to the qubit's oscillating E-field. When TLS are at resonance with the qubit, they can efficiently dissipate energy into the phonon~\cite{Lisenfeld:PRL:2010} or BCS-quasiparticle~\cite{Bilmes17} bath which results in reduced qubit energy relaxation times $T_1$~\cite{Barends13} and, in the case of strong coupling, gives rise to avoided level crossings in qubit spectroscopy~\cite{Simmonds:PRL:2004}.
Moreover, thermally activated TLS at low energies may interact with high-energy TLS that have frequencies near resonance with a qubit or resonator, and this causes temporal fluctuations of the device's resonance frequency~\cite{FaoroResonators,schlor2019} and energy relaxation rate~\cite{Faoro:PRL:2012,klimov2018,burnett2019,Mueller:2014}.\\

Further progress with superconducting quantum processors based on current circuit architectures thus requires extensive material and fabrication process research to avoid the formation of TLS. Moreover, tools to verify the quality of metal films and junctions are required that are able to relate fabrication processes to TLS formation and to investigate the microscopic nature of the material defects. For these tasks, qubits themselves are well-suited because of their sensitivity to TLS. In case of strong coupling, quantum state swapping between the qubit and TLS~\cite{Cooper04} is possible, allowing one to characterize TLS' coherence properties~\cite{Neeley2008, Shalibo:PRL:2010,Lisenfeld:PRL:2010}, and their coupling to the environment~\cite{Lisenfeld2016,Bilmes17,Meissner18}. A useful method for such studies is to control the TLS' internal asymmetry energy and thus their resonance frequency by applied mechanical strain~\cite{Grabovskij12} or DC-electric field~\cite{Sarabi16}. Operating qubits in electric fields allows one to distinguish defects in tunnel junction barriers from those on electrode surfaces~\cite{Lisenfeld19} and to obtain information on the positions of individual TLS in the quantum circuit~\cite{Bilmes20}.\\

In this letter, we present a quantum sensor that grants access to measurement and manipulation of individual TLS in virtually arbitrary materials. The device is based on a transmon qubit~\cite{KochTransmon,Barends13} which consists of a capacitively shunted DC-SQUID formed by two Josephson junctions connected in parallel, as shown in Fig.~\ref{fig:fig1}\,\textbf{a} and\,\textbf{c}. Qubit readout is performed by measuring the dispersive resonance frequency shift of a coplanar resonator that is capacitively coupled to the qubit. The qubit resonance frequency can be tuned in a range of typically $\approx$~1~GHz by an on-chip coil providing magnetic flux which frustrates the Josephson energy of the DC-SQUID loop shown in Fig.~\ref{fig:fig1}\,\textbf{d}.\\

The material under test defines the dielectric in an additional ``sample capacitor'' shunting the transmon qubit. In this work, we use a capacitor having a plate or "overlap" geometry as shown in Fig.~\ref{fig:fig1}\,\textbf{e}.
This allows one to study TLS in all dielectrics that can be deposited as thin-films, e.g. by sputtering or evaporation.
Alternatively, one can employ a so-called nanogap capacitor consisting of two coplanar electrodes (see Fig.~\ref{fig:fig1} \textbf{f}) that are separated by a few tens of nanometres, and then covered by the sample material. In this case, the coupling between TLS and the qubit occurs via the fringing electric field sketched in the inset of Fig.~\ref{fig:fig1} \textbf{f}.
This provides the possibility to study TLS in pieces of bulk material by pressing it onto the nanogap capacitor.
Moreover, the use of uncovered nanogap capacitors allows one to study single TLS residing in the native oxides of the electrode material and defects that are formed by surface adsorbates.\\

The STM describes TLS on the basis of a double-well potential whose minima differ by the asymmetry energy $\varepsilon$, and transitions between wells occur at a tunneling energy $\Delta_0$, resulting in the transition energy $E=\sqrt{\Delta_0^2 + \varepsilon^2}$.
TLS in the sample material couple to the qubit at a strength $\hbar g = \mathbf{pF} = \bar{p} |\mathbf{F}|$, where $\mathbf{F}$ is the electric field inside the capacitor which is induced by the qubit plasma oscillation, and $\bar{p} = p_\parallel\, (\Delta_0/E)$ is the projection of the TLS' dipole moment $\mathbf{p}$ onto $\mathbf{F}$, multiplied by the TLS' matrix element~\cite{Sarabi16} $\Delta_0/E$.\\

\subsection*{Results} \noindent

\begin{figure*}[htbp]
	\begin{center}
	\includegraphics[width=\textwidth]{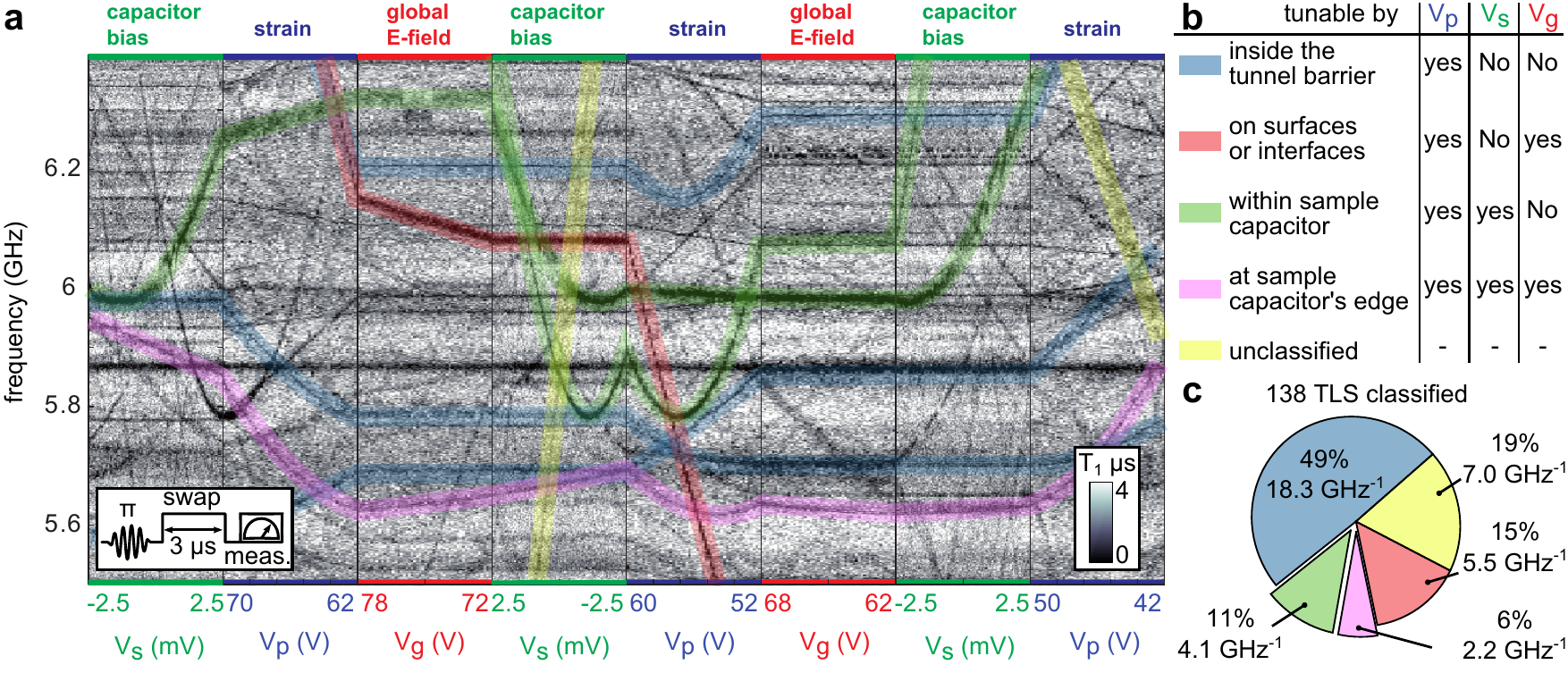}			
	\end{center}
	\caption{
	\textbf{Defect spectroscopy.}
	\textbf{a} Typical data recorded with the swap spectroscopy protocol (see left inset). Each dark trace indicates a reduction in the qubit's $T_1$ time due to resonant coupling to an individual TLS. The response of TLS to control parameters (such as physical strain, global E-field, and the E-field inside the sample capacitor) allows one to identify the possible location of the TLS as listed in the table \textbf{b}. 
	Some exemplary TLS are highlighted with colored lines, where yellow lines indicate TLS that could not be classified since they were visible in only a single segment. 
	While $\vg$ and $\vp$ were continually reduced, $V_\text{s}$ was ramped alternatingly up or down with the amplitude limited to $|V_\text{s}| < 2.5$mV to avoid heating of the attenuators in the bias line.
	\textbf{c} The resulting density of TLS (detected TLS per GHz bandwidth) of 138 TLS observed in measurements on two identical qubits. 	
}
	\label{fig:swapspec}
\end{figure*}

Single TLS can be detected if their energy exchange rate with the qubit (which equals their coupling strength $g$ at resonance) is comparable to the energy decay rate $1/T_1$ of the isolated qubit. 
The criterium $g \approx 1/T_1$ togeher with the TLS' above-mentioned coupling strength $\hbar g = \bar{p} |\mathbf{F}|$ define a suitable thickness $d$ of the dielectric layer in overlap capacitors: $d = \bar{p}\,T_1\,  V_\mathrm{rms} /\hbar$, where 
the electric field $|\mathbf{F}|$ is substituted by $V_\mathrm{rms}/d$. Here, $V_\mathrm{rms}= \sqrt{\hbar \omega_{10} /2\ctot}\approx 4.5\,\micro\mathrm{V}$ is the vacuum voltage fluctuation on the qubit island at the designed plasma oscillation frequency $\omega_{10}\approx 2\pi\cdot 6.2\,\mathrm{GHz}$ when $\ctot \approx 100\, \mathrm{fF}$ is the sum of all capacitances shunting the qubit.
To be able to detect a TLS dipole moment $p_\parallel$ of minimum $0.1\, e$\AA~\cite{Martinis:PRL:2005}, and assuming a rather conservative $T_1 \approx 1\micro\second$, we arrive at a dielectric layer thickness $d \approx 70\,\mathrm{nm}$. We chose $d=50\,\mathrm{nm}$ and a capacitor size of $(0.25\cdot 0.30)\, \micro\meter^2$, resulting in $C_s \approx 0.15$ fF $\ll \ctot$ which ensures that the energy that is stored in the lossy sample capacitor remains limited to a small fraction of the qubit's total energy, and coherence is preserved. A picture of the employed sample capacitor is shown in Fig.~\ref{fig:fig1}\,\textbf{e} while further details on the capacitor design are given in Supplementary Methods 1.\\

It is furthermore necessary to be able to distinguish TLS in the sample material from those on electrode interfaces and from TLS in Josephson junctions. This is accomplished by probing the TLS' response to a local electric field generated by voltage-biasing the sample capacitor's electrode as indicated in Fig.~\ref{fig:fig1}\textbf{a}, where the additional capacitor \cf$\sim 250\,\mathrm{fF}$ serves as a DC-break.
The bias voltage will not induce an electric field in the transmon's shunt capacitor $\ctrans$ nor inside the Josephson junctions' tunnel barrier since the DC-electric potential difference of the transmon island and ground will be compensated by Cooper pair tunneling~\cite{Lisenfeld19}, so that only TLS in the sample capacitor respond to the applied voltage $\vs$. In addition, we can tune TLS residing at the perimeter of the qubit capacitor by a globally applied DC-electric field that is generated by an electrode installed above the qubit chip~\cite{Lisenfeld19} as shown in Fig.~\ref{fig:fig1}\,\textbf{b}. Moreover, all TLS including those residing inside the tunnel barriers of junctions can be tuned via physical strain by bending the chip with a piezo actuator~\cite{Grabovskij12,Lisenfeld2015} which is useful to enhance the number of observable TLS. The table in Fig.~\ref{fig:swapspec} \textbf{b} summarizes how to identify the location of a TLS from its tunability characteristics.\\

We chose amorphous aluminum oxide AlO$_x$ as the sample material for this work since it is well characterized and of general importance for superconducting quantum circuits where it is ubiquitously used as a reliable tunnel barrier material.
The sample capacitor is patterned with electron-beam lithography, where the bottom electrode is deposited and connected to the qubit island in the same step as the qubit's Josephson junctions, followed by a third lithography step depositing 50nm of AlOx by eBeam-evaporation of Al in an oxygen atmosphere, and capping it by a top Al electrode. The filter capacitor $\cf$ is formed simultaneously as a wider section in the top electrode. 
Here, we report results for samples employing small sample capacitors of size $(0.25\micro\meter\times0.30\micro\meter)$ which did not contribute significantly to decoherence. Two tested $C_s$-shunted qubits reached $T_1$-times of 3.3 to 4.2 $\micro \second$, which is comparable with an isolated reference qubit ($T_1 \approx 4.3 \micro \second$) on the same chip. In another batch, we also tested larger sample capacitors $(0.3\micro\meter\times2.1\micro\meter)$ which did limit the qubit's $T_1$ time~\cite{Bilmes19}. This allowed us to measure the loss tangent of the employed AlO$_x$ dielectric as $\tan \delta_0 \approx (1.7\pm0.2)\cdot 10^{-3}$, comparable with other reports~\cite{Martinis:PRL:2005,Pappas2011,Chunging2014,Brehm2017}.\\

To distinguish whether a TLS is located in a tunnel barrier, at the qubit's film edges~\cite{Bilmes20}, or in the sample capacitor dielectric, we track its resonance frequency for a range of voltages applied to the global DC electrode ($V_\mathrm{g}$), to the sample dielectric ($\vs$), and to the piezo ($\vp$). An example of such a measurement is presented in Fig.~\ref{fig:swapspec} \textbf{a}, showing the frequency dependence of the qubit's $T_1$ time estimated by swap spectroscopy~\cite{Lisenfeld2015,Cooper04,Shalibo:PRL:2010}, where dark traces reveal enhanced qubit energy relaxation due to resonant TLS. These segmented hyperbolic traces are fitted to obtain the TLS'  coupling constants $\gamma$ which determine their bias-dependent asymmetry energy $\varepsilon = \varepsilon_\text{i} + \gamdc \vg + \gams \vs + \gamp \vp$ up to an intrinsic offset $\varepsilon_\text{i}$. The fit also results in the value of the TLS' tunneling energy $\Delta_0$ if it lies within the tunability range of the qubit's resonance frequency.\\

Thanks to the well-specified DC-electric field $\vs/d$ in the sample capacitor, the coupling electric dipole moment $p_\parallel = \gams d/2$ of TLS in the sample material is directly obtained from the identity $2p_\parallel\vs/d=\gams\vs$~\cite{Anderson:PhilMag:1972,Phillips:JLTP:1972}. In contrast, a measurement of the TLS' coupling strength to a quantum circuit results in the effective dipole moment size $\bar{p}$ where the matrix element $(\Delta_0/E)$ is often unknown. From measurements on two identical qubits in one cool-down, we characterized in total 138 TLS. Of those, 13 TLS were found inside the sample material, with a spectral density of $4.1\,\mathrm{GHz}^{-1}$ (see calculation details in Supplementary Methods 3) which results in a volume density of $P_0 = 4.1 \,(V_\text{d} \,\mathrm{GHz})^{-1} = 1800\,(\micro \mathrm{m}^3\cdot \mathrm{GHz})^{-1}$. We estimated the field-free dielectric volume $V_\text{d}=(0.15\cdot 0.3 \cdot 0.05)\, \micro\meter^3$ by assuming that the global field penetrates the sample dielectric to a depth of about its thickness ($50\,\nano\meter$) from the sides open to air.\\

\subsection*{Discussion} \noindent

 The average dipole moment of the observed sample-TLS was $p_\parallel=(0.4\pm0.2)\,\mathrm{e}$\AA~(see Supplementary Methods 2 for calculations), which results in a loss tangent~\cite{Gao08} of the employed AlO$_x$ ($\varepsilon_r \approx 10$) of $\tan \delta_0 = \pi P_0 p_\parallel^2 (3 \varepsilon_0 \varepsilon_r)^{-1} \approx 1.0\cdot 10^{-3}$, comparable with the number quoted above.
The statistics shown in Fig.~\ref{fig:swapspec} \textbf{c} indicate that the qubits were mostly limited by TLS hosted inside the tunnel barrier of the stray Josephson junctions (light green in Fig.~\ref{fig:fig1} \textbf{d}) which are a fabrication artefact that could have been avoided by shorting them in an  additional lithography step~\cite{quintana2014,Bylander2020}.\\

For the $1.5$ to $2\,\nano\meter$-thin~\cite{Kang2014,Zeng2015,Fritz2019} and $17.17\,\micro\meter^2$-large tunnel barriers of the two stray junctions shown in Fig~\ref{fig:fig1} \textbf{d}, our measurements indicate a TLS volume density of $P_{0,\text{JJ}} = 360$ to $270\,(\micro \mathrm{m}^3\cdot \mathrm{GHz})^{-1}$, in good agreement to previous work~\cite{Lisenfeld19}. Notably, this is about six times smaller than the TLS density found in the thicker layer of deposited AlOx in the sample capacitor. This is probably due to the minimum detectable TLS dipole moment size, i.e. qubit's sensitivity, which is smaller for sample-TLS due to stronger oscillating qubit fields ($\approx 90\,\mathrm{Vm^{-1}}$) inside the sample capacitor, compared to the field inside the tunnel barrier of the stray junctions ($\approx 15\,\mathrm{Vm^{-1}}$). We speculate that this notion might be dressed due to various reasons like a reduced dangling bond density due to facilitated atom diffusibility and self-annealing in the thin tunnel barrier~\cite{Molina-Ruiz2020}, or enhanced shielding of TLS by the evanescent Cooper-pair condensate~\cite{Bilmes19}, or reasons related to the material’s different growth conditions.\\

\begin{figure}[htbp]
		\includegraphics[width=\columnwidth]{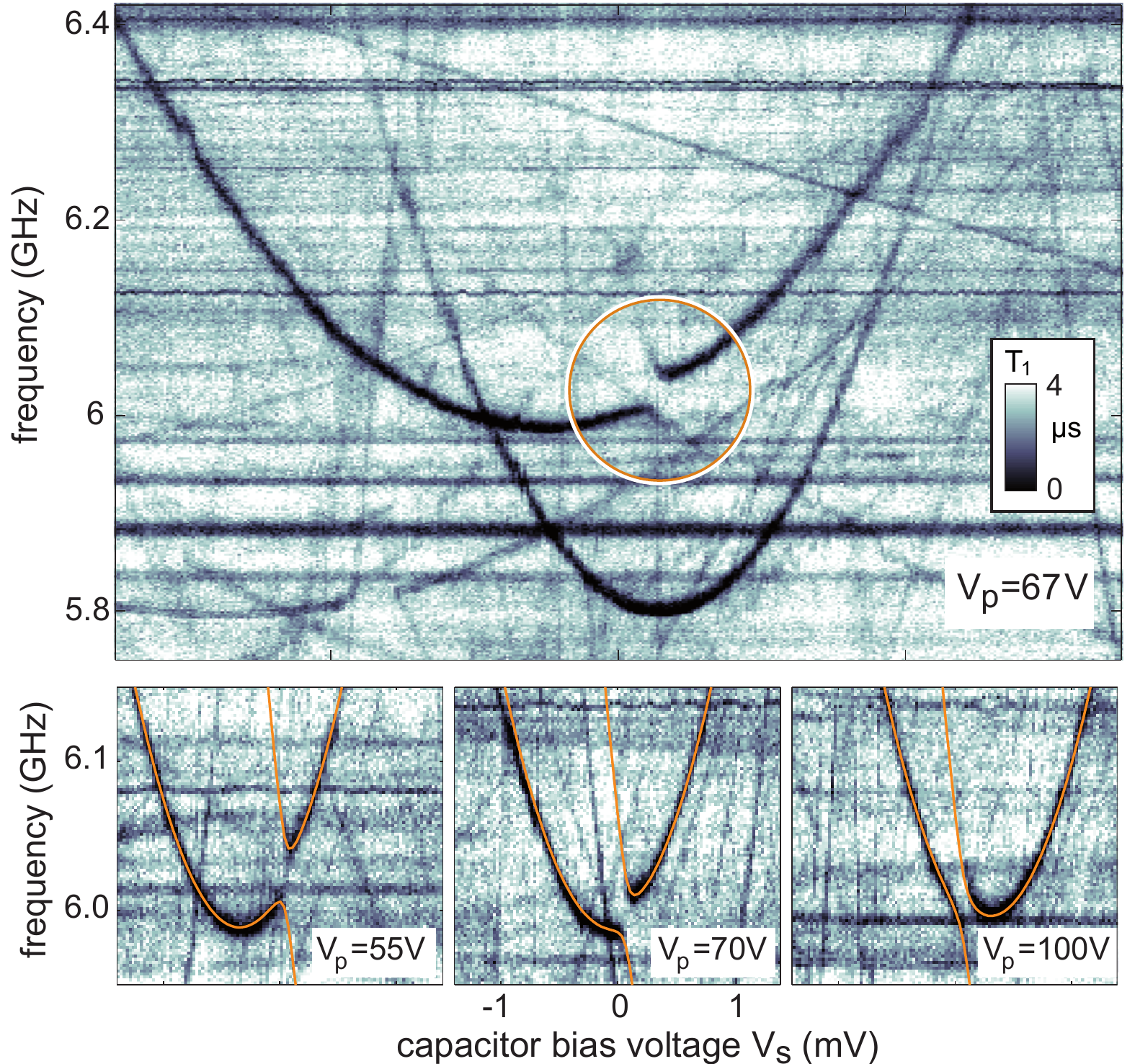}
		\caption{
		\textbf{Interacting TLS in the sample dielectric.}
		\textbf{Top:} Avoided level crossing (encircled) in the spectrum of a TLS due to coherent interaction with a second TLS. \textbf{Bottom:} The observed level splitting could be shifted through the TLS symmetry point by mutually detuning the two TLS via the physical strain. Each figure was recorded for the same range of bias voltages but at different voltage $\vp$ applied at the piezo actuator. Superimposed orange lines show the transition frequencies calculated using independently measured TLS parameters and best-fitting interaction strengths.}
	\label{fig:interactions}
\end{figure}

E-field spectroscopy also revealed coherent mutual interactions between TLS in the sample material which manifest themselves in avoided level crossings as shown in Fig.~\ref{fig:interactions}.
The coupling between the TLS is described by the interaction Hamiltonian $H_\mathrm{int} = \frac{\hbar}{2} (g_x \sigma^x_{1} \sigma^x_{2} + g_z \sigma^z_{1} \sigma^z_{2})$, where $ \sigma^x_{i}$ and $\sigma^z_{i}$ are the Pauli matrices of TLS $i$. 
As an advancement over earlier work~\cite{Lisenfeld2015}, the combined control of strain and local E-field allowed us to mutually detune the TLS and shift the avoided level crossing through the symmetry point of the observed TLS as demonstrated by the lower panels of Fig.~\ref{fig:interactions}. Since the longitudinal coupling component $g_z \propto \sigma^z_{1}$ changes its sign when TLS 1 is tuned through its symmetry point, its effect can be well distinguished from the transversal component $g_x$. 
This enabled fitting of both components $g_x = -19\,(\micro \mathrm{s})^{-1}$ and $g_z = 25\,(\micro \mathrm{s})^{-1}$. More details on the description of coherently interacting TLS can be found in a previous work~\cite{Lisenfeld2015} and in Supplementary Methods 4.\\

In conclusion, we demonstrated that superconducting qubits can serve as interfaces for studying quantum properties of individual atomic-size tunneling systems located in arbitrary materials deposited as thin films. Qubit swap spectroscopy in dependence on the applied electric field bias to the sample material enables precise measurement of the TLS' coupling dipole moments and reveals avoided level crossings which herald coherent interaction between TLS. The possibility to mutually detune interacting TLS by using mechanical strain as a second control parameter allows one to fully characterize the type of the interaction.
The demonstrated approach has a large potential to provide further insights into the puzzling physics of amorphous solids. It may serve as a valuable tool in the search for low-loss materials urgently needed to advance nano-fabricated devices and superconducting quantum processors where TLS play a detrimental role.

\subsection*{Methods} \noindent

\noindent\textbf{Sample fabrication}\\
The qubit samples were fabricated and characterized at KIT. A microchip contained three independent Xmon qubits of whom two were shunted by a sample capacitor, and a third one served as a reference qubit. The qubit electrode, ground plane and resonators were patterned into a $100\,\mathrm{nm}$ thick Al film with an inductively coupled Ar-Cl plasma. After Argon-ion milling~\cite{Gruenhaupt2017} of the optically patterned electrodes in a PLASSYS shadow evaporation device, the Josephson junctions were deposited in a subsequent electron-beam lithography step.\\

Qubit samples with large and small sample capacitors were studied. The bottom electrode of large sample capacitors consisted a narrow extension of the qubit island. The bottom layer of the small sample capacitor (see Fig.~\ref{fig:fig1} \textbf{e}) was made simultaneously with the Josephson junctions. Sample dielectric and top electrodes of both capacitor types were formed in the PLASSYS device using an MMA/PMMA copolymer mask patterned in an electron-beam lithography step. After removing the native oxide of the bottom electrode with the Ar milling process, the sample dielectric (here $50\,\mathrm{nm}$ AlO$_x$) was formed during a perpendicular deposition of Al at a rate of $0.2\,\mathrm{nm\,s}^{-1}$ in an oxygen atmosphere (chamber pressure of $3\times10^{-4}\,\mathrm{mBar}$, oxygen flow of $5\,\mathrm{sccm}$). The dielectric was in-situ covered by perpendicularly deposited $100\,\mathrm{nm}$-thick layer of Al that formed the top electrode. Further details are reported in the PhD thesis by AB, Chap. 3.2.3~\cite{Bilmes19}.\\

\noindent\textbf{Experimental setup}\\
The sample was measured in an Oxford Kelvinox 100 wet dilution refrigerator at a temperature of 30 mK. The qubit chip was installed in a light-tight aluminium housing protected by a cryoperm magnetic shield. The coaxial control lines were heavily attenuated, filtered, and equipped with custom-made infrared filters. The qubit state was detected via the dispersive shift~\cite{wallraff2004} of a notch-type readout resonator which was capacitively coupled to the qubit, and probed in a standard homodyne microwave detection setup.\\

The DC-gate for tuning the surface-defects consisted of a copper-foil/Kapton foil stack that was glued to the lid of the sample box. It was connected via a twisted pair equipped with an RC-lowpass filter (cutoff ca 10 kHz) at the 1K-stage, and a custom-made copper powder lowpass filter (1 MHz cutoff) at the 30 mK stage. The top electrode of the sample capacitor was controlled via an attenuated microwave line, as further detailed in the Supplementary Methods 2.\\


\subsection*{Acknowledgements}\noindent
AB acknowledges support from the Helmholtz International Research School for Teratronics (HIRST) and the Landesgraduiertenförderung-Karlsruhe (LGF). JB was financially supported by Studienstiftung des Deutschen Volkes. JL gratefully acknowledges funding from the Deutsche Forschungsgemeinschaft (DFG), grant LI2446-1/2. AVU acknowledges support  provided by the Ministry of Education and Science of the Russian Federation in the framework of the Program to Increase Competitiveness of the NUST MISIS (contract No. K2-2020-017). The work was also supported by the Initiative and Networking Fund of the Helmholtz Association and by Google LLC. We acknowledge support by the KIT-Publication Fund of the Karlsruhe Institute of Technology. We acknowledge Johannes Rotzinger and Ioan Pop for fruitful discussions, Silvia Diewald and Patrice Brenner for their technical assistance with electron-beam devices, and Lucas Radtke for his indispensable assistance in the clean-room.

\subsection*{Author contributions}\noindent
The qubit samples were designed and fabricated by AB. Experiments were devised and performed by JL in a setup implemented by AB and JL. 
SV performed calculations for the mutually coupled TLS system, and JB simulated the electric-field distribution of nanogap capacitors. The manuscript was written by JL and AB with contributions from all authors. 


\subsection*{References}\noindent

\clearpage
\onecolumngrid
\begin{center}
\large
\textbf{Supplementary materials for\\Quantum Sensors for Microscopic Tunneling Systems\\}
\vspace{0.25cm}
\textbf{Supplementary Material\\}
\date{\today}
\normalsize
\vspace{0.25cm}
Alexander Bilmes, Serhii Volosheniuk, Jan David Brehm, Alexey V. Ustinov, and J\"urgen Lisenfeld\\
\end{center}

\section*{Supplementary Methods}
\subsection*{Sample capacitor design}
\label{ss_OVsize}
The choice of the sample capacitor dimensions (in plate geometry) is a trade-off between the participation ratio of the sample dielectric which influences the qubit coherence, and the dielectric thickness $d$ which determines the typical coupling strength of the qubit to sample TLS. As explained in the main text, we chose $d = 50$ nm to balance the TLS-qubit coupling strength $g$ against an expected qubit energy relaxation time of $T_1 \approx 1\,\micro s$. Further, to determine the maximum sample capacitor area $A$, we consider the qubit's energy relaxation rate expressed in terms of dielectric losses contributed by the overlap capacitor:
\begin{align}
\Gamma_1&=2\pi f_{01}\cdot p_\text{s}\cdot \tan\delta_0+\Gamma_{1,0}.\label{eq_partrat}
\end{align}
Here, $\tan\delta_0$ is the loss tangent of the sample dielectric, and $\Gamma_{1,0}$ combines all other sources of loss. The sample capacitor's participation ratio $p_\text{s}=\csamp/(\csamp+C)$ is defined as the electric energy stored in the sample capacitor divided by the total energy of the qubit. Here, $\csamp=\varepsilon_0\varepsilon_\text{r}A/d$ is the sample capacitance, and $C$ is the sum of all other capacitors shunting the qubit's Josephson junctions. Assuming an energy relaxation rate $\sim 1/(10\,\micro\mathrm{s})$ of the qubit without a sample capacitor, and the loss tangent of $1.6\times10^{-3}$ in AlO$_x$~\cite{Martinis:PRL:2005}, we arrived at a sample capacitor size of ($300\,\mathrm{nm}\times2.1\,\micro\mathrm{m}$) used in the first generation of samples. As reported in Chap.~4.1 of AB's thesis ~\cite{Bilmes19}, these samples were used to deduce a loss tangent of 
$\tan\delta_0=1.7\times10^{-3}$ of the employed aluminum oxide. To improve the signal-to-noise ratio for TLS spectroscopy, the sample capacitor size ($250\,\mathrm{nm}\times 300\,\mathrm{nm}$) of the second sample generation was designed for a reduced $\Gamma_1$ of $1/(3\,\micro\mathrm{s})$, following Eq.~\eqref{eq_partrat} and values of $\Gamma_{1,0} \sim 1/(5\,\micro\mathrm{s})$ and $\tan\delta_0$ measured with the first generation samples.\\

\subsection*{Measuring the electric dipole moments of sample TLS}
\label{ss_dipole}
As mentioned in the main text, the sample TLS are identified by their tunability via the DC-voltage $\vs$ maintained across the sample capacitor, while they do not respond to the globally applied DC-electric field generated by the gate placed above the microchip. As a note, the simulated distribution of the global field~\cite{Bilmes19} shows that all detectable TLS which reside at qubit circuit interfaces other than in the Josephson junction or in the sample capacitor, are exposed to the global field. Moreover, the TLS hosted in the qubit's Josephson contacts are not tunable by $\vs$ since the qubit's island has a constant DC-electric potential due to the transmon regime.\\

\begin{figure}[htbp]
	\begin{center}
		\includegraphics[width=.45\columnwidth]{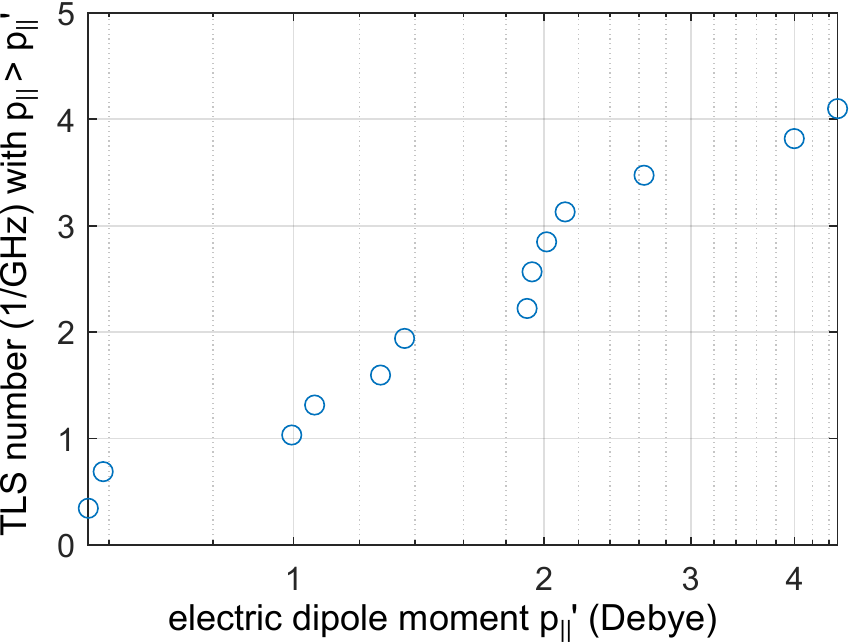}
	\end{center}
	\caption{Electric dipole moment sizes of TLS hosted in the AlO$_x$ sample dielectric. Data comprises measurements on two qubits.}
	\label{fig:SM_dipoles}
\end{figure}

From a fit to the hyperbolic dependence of the TLS' resonance frequency $f=\sqrt{\Delta_0^2 + (\varepsilon_\text{i} + \gams \vs)^2}/\mathrm{h}$ on the applied voltage $\vs$, we extract the tunability coefficient $\gams$ which is related to the TLS' electric dipole moment component $p_\parallel$ parallel to the field by $\gams \vs\equiv 2p_\parallel \vs/d $. The latter term corresponds to the dipole energy in the electric field $\vs/d$ which adds to the intrinsic asymmetry energy $\varepsilon_\text{i}$. The cumulative spectral density of sample TLS, which is plotted vs. the deduced dipole moment size in Fig.~\ref{fig:SM_dipoles}, is comparable to previous reports~\cite{Martinis:PRL:2005,Sarabi16,Brehm2017,Bilmes19}.\\

The voltage $\vs$ was controlled via an attenuated coaxial RF line whose warm end was connected to a voltage source (output voltage $\widetilde V_\text{s}$), and the cold end was connected to the sample capacitor's top electrode. This RF line consisted a chain of following elements: a $-20\,\mathrm{dB}$ attenuator at room temperature, $\sim 50\,\mathrm{cm}$ stainless-steel (SS) coaxial cable leading to the $4\,\mathrm{K}$ plate, two $-10\,\mathrm{dB}$ attenuators at $4\,\mathrm{K}$, further $\sim 30\,\mathrm{cm}$ of SS coax leading to a $-10\,\mathrm{dB}$ at the still plate, and another $\sim 15\,\mathrm{cm}$ Cu-coax leading to a $-3\,\mathrm{dB}$ attenuator at the mixing chamber. We have measured at ambient conditions the voltage drop $\vs$ between the $-3\,\mathrm{dB}$ attenuator's central pin and ground as function of $\widetilde V_\text{s}$, and found a division factor of $\widetilde V_\text{s}/\vs=205$.\\ 

\subsection*{Calculating the TLS spectral density}
\label{ss_density}
Here, we qualitatively describe our method to deduce the spectral TLS density from data sets such as that shown in the figure~2 \textbf{a} of the main text. A quantitative description can be found in chapter 4.2.4 of the PhD thesis by AB~\cite{Bilmes19}.\\

The data set of TLS spectroscopy, such as that shown in Fig.~2 \textbf{a}, shows estimates of the qubit's $T_1$-time (see color bar) as a function of its resonance frequency (vertical axis). Each trace along frequency is obtained using the swap spectroscopy protocol shown in the inset of \textbf{a} while one of the control voltages (piezo, DC-electrode or sample capacitor) is swept (horizontal axes). The TLS spectral density is the average number of TLS detectable per one such data trace.
For each observed TLS, we calculate its spectral density by relating the control voltage range in which its trace is observed to the whole voltage range.
The spectral density of a given class of TLS shown in \textbf{b} then is the sum over the spectral densities of individual TLSs which belong to this class.\\

As an example, take the data set shown in Fig.~2 \textbf{a}. If a TLS appears in all segments (plot regions with green, blue or red horizontal axes), then its spectral density is one over the observed frequency range, thus $1/(0.9\,\mathrm{GHz})$, as in the case of the violet-colored TLS trace. As an opposite example, the yellow highlighted TLS trace in the fourth segment from left is visible in only $50\%$ of the data traces in this segment. Thus, its spectral density is $0.5/8/(0.9\,\mathrm{GHz})$ where $0.5/8$ is the relative part of all eight segments where the TLS is visible. The red colored trace partially appears in four segments, and accordingly the spectral density of this TLS is $(0.3+1+1+0.7)/8/(0.9\,\mathrm{GHz})$.\\

\subsection*{Characterization of interacting TLS}
\label{ss_interaction}

\noindent Here, we describe the model of interacting TLS used to fit the data shown in Fig.~3 of the main text. A more detailed description of the same model is given in Lisenfeld et al.~\cite{Lisenfeld2015} and its supplementary material.  \\

\noindent The Hamiltonian of a single TLS is written as 
\begin{equation}
H_{i} = \frac 12 \varepsilon_i(V_\mathrm{p},\vs,\vg)\sigma_{\text{z},i} + \frac 12 {\Delta}_i\sigma_{\text{x},i} = \frac 12 E_i(V_\mathrm{p},\vs,\vg)\,	\widetilde{\sigma}_{\text{z},i},
\label{eq:H_TLS}
\end{equation}
where ${\Delta}_i$ is the tunneling energy and the asymmetry energy $\varepsilon_i(V_\mathrm{p},\vs,\vg) = \varepsilon_{i,0} + \gamma_{\text{p},i} V_p + \gamma_{\text{s},i} \vs + \gamma_{\text{g},i} \vg$ depends linearly on external strain (set by the piezo voltage $V_\mathrm{p}$), on the local electric field (controlled by the voltage $\vs$), and on the global electric field (set by $\vg$). Here, we use the tilde to distinguish operators such as the Pauli-matrices  $\widetilde{\sigma}$ in the eigenbasis from those in the localized basis $\sigma$ that is spanned by the two positions of the tunneling entity.
The energy splitting in the diagonal basis is $E_{i}(V_\mathrm{p},\vs,\vg) = \sqrt{\varepsilon_{i}^{2}(V_\mathrm{p},\vs,\vg) + {\Delta}_{i}^{2}}$. \\



\noindent The interacting TLS system is described by the interaction Hamiltonian $H_{\text{T}}=H_1+H_2+H_{12}$,
with the coupling terms
\begin{equation}
H_{12} = \,g\,\sigma_{\text{z},1}\,\sigma_{\text{z},2}/2 .
\label{eq:H_12}
\end{equation}


\noindent In the diagonal basis, the interaction Eq.~(\ref{eq:H_12}) consists of four terms, which are combinations of $\widetilde{\sigma}_{\text{z}}$ and $\widetilde{\sigma}_{\text{x}}$ for each TLS. The resulting Hamiltonian, $\widetilde{H}$, can be significantly simplified~\cite{Lisenfeld2015} by neglecting all coupling terms of the form $\propto \widetilde{\sigma}_x\widetilde{\sigma}_z$ and $\propto \widetilde{\sigma}_z\widetilde{\sigma}_x$. 
They represent a coupling where one partner changes its state depending on the instantaneous state of the other subsystem and contribute only as small energy offsets.
The significant parts of the interaction Hamiltonian using the longitudinal and transversal coupling factors $g_{\text{z}}$ and $g_{\text{x}}$ are thus
\begin{equation}
\widetilde{H}_{12}= \frac{1}{2} (g_\text{z}\widetilde{\sigma}_{z,1}\widetilde{\sigma}_{z,2}+g_\text{x}\widetilde{\sigma}_{x,1}\widetilde{\sigma}_{x,2}).
\end{equation}

\begin{figure*}[h!]
	\begin{center}
		\includegraphics[width=1\columnwidth]{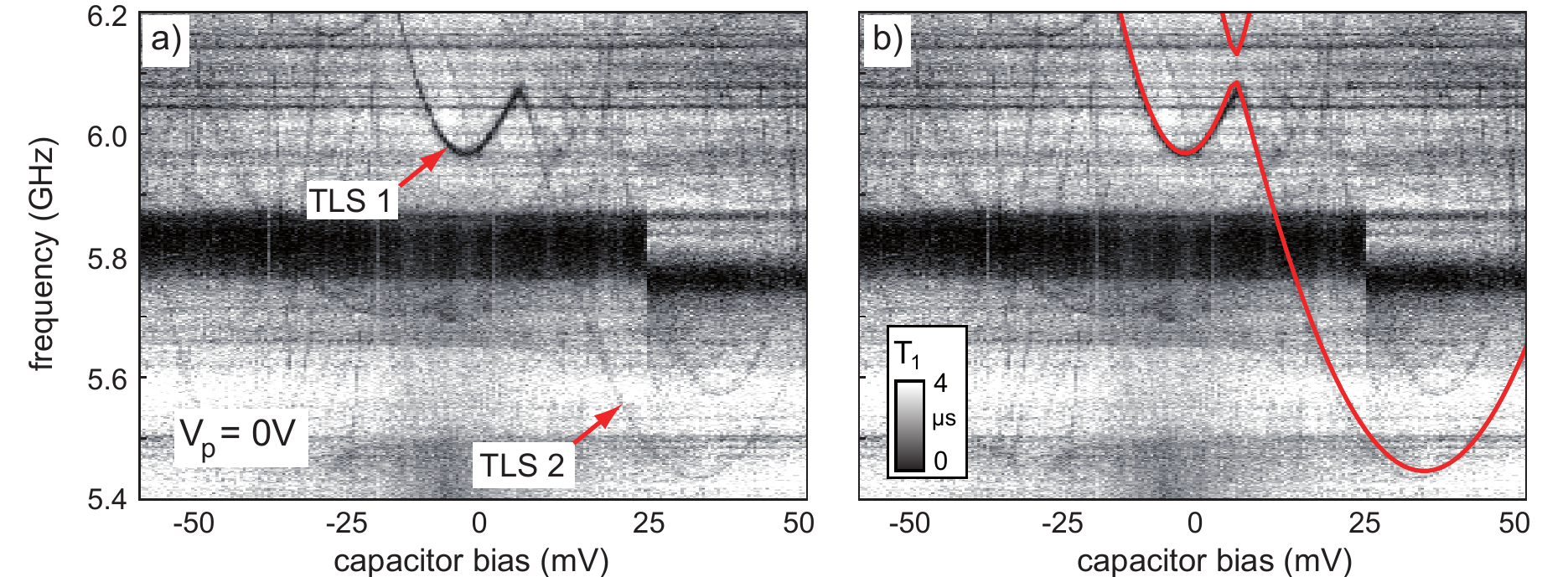}
	\end{center}
	\caption{Swap-spectroscopy of the interacting TLS system in a wider frequency range, obtained at $\vp = 0$V. \textbf{a)} Raw data, where the arrows indicate the traces of TLS 1 and TLS 2. \textbf{b)} Same data with superimposed fits calculated with the extracted TLS parameters shown in Table~\ref{tabresults}.}
	\label{figs:interactions}
\end{figure*}

\noindent To characterize the interacting system, we first measure the TLS' tunneling energies $\Delta_i$ and the dependence of their asymmetry energies $\varepsilon_i$ on the applied local E-field using swap-spectroscopy acquired in a wider frequency range as shown in Fig.~\ref{figs:interactions}. Next, the coupling strength of TLS 1 to the applied mechanical strain $\gamma_{\text{p},1}$ is obtained from swap-spectroscopy scans taken at different strain values using data as shown in Fig.~3 of the main text. We obtain the remaining parameters $\gamma_{\text{p},2}$, $g_\text{z}$, and $g_\text{x}$ by fitting these data sets to a quantum simulation of the system Hamiltonian implemented with the QuTip software package~\cite{qutip}. Table~\ref{tabresults} summarizes the resulting TLS parameters. The obtained coupling strengths are 
$g_\text{z} \approx 25.0$ MHz, and $g_\text{x} \approx -19.0$ MHz.\\

\begin{table}[h!]
\centering
 \begin{tabular}{|lll|l|l|}
\hline
\textbf{parameter} &&& \textbf{TLS} 1 & \textbf{TLS 2}\\[1ex]
 \hline
tunnel energy & $\Delta_i$ & (GHz) & 5.957 &  5.440\\
local field coupling&  $\gamma_{\text{s},i}$ &(GHz/\vs)& 161.95 & 92.25 \\
el. dipole moment & $p_\parallel$ &(e\AA)& 0.335 & 0.191\\
strain coupling&  $\gamma_{\text{p},i}$& (MHz/\vp) & 22 & 0 \\
\hline
\end{tabular}
 \caption{Parameters of the interacting TLS, obtained by fitting the data shown in Fig. 3 of the main text and the data shown in Fig.~\ref{figs:interactions}. For TLS 2, no strain dependence could be detected.}
 \label{tabresults}
\end{table}

\subsection*{Deposition of alignment marks}

To minimize contamination and manufacturing time, the niobium alignment marks, which were required for the electron-beam lithography, were deposited on the same resist mask that was used to etch the main structures into the Al film. During Nb deposition, the main structures were covered with a protecting wafer as further detailed in the PhD thesis by AB, Chap. 3.2.1~\cite{Bilmes19}.\\

\section*{Supplementary Figures}
\label{ss_data}
\begin{figure*}[htbp]
	\begin{center}
	\includegraphics[scale=1]{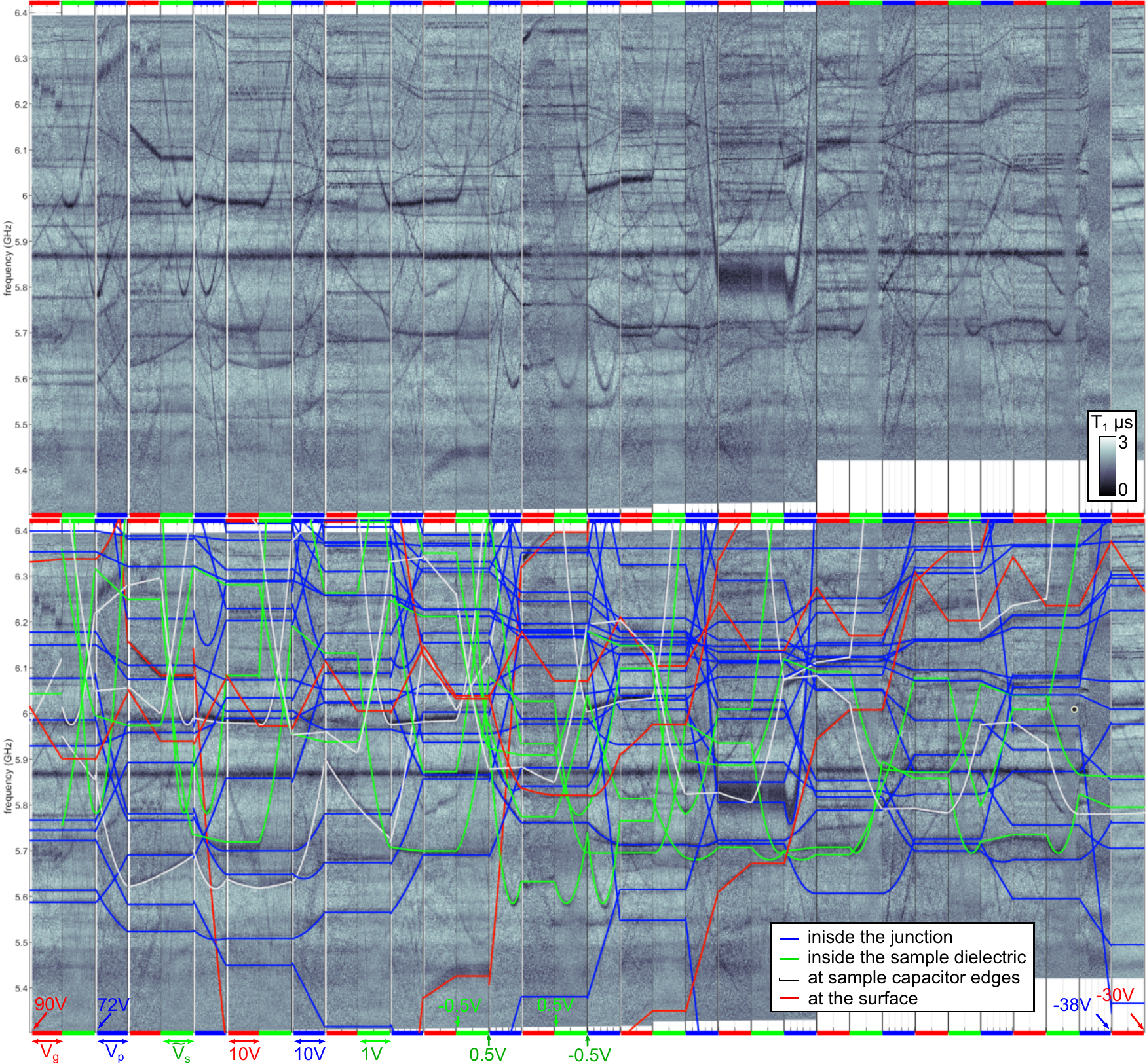}		
	\end{center}
	\caption{Swap-spectroscopy data to measure the response of TLS to the global DC-electric field controlled by  $\vg$ (red framed segments), to the local field applied directly to the capacitor electrode via the voltage $\widetilde V_\text{s}$ (green segments), and to mechanical strain controlled by the piezo voltage  $\vp$ (blue). Top panel: raw data, bottom panel: with superimposed hyperbolic fits to the TLS' resonance frequencies. While $\vg$ and $\vp$ were both ramped linearly down from about 90V to -30V, $\widetilde V_\text{s}$ was ramped alternatingly up or down with the amplitude limited to  $|\widetilde V_\text{s}| < 0.5$V to avoid heating of the attenuators. Data recorded on qubit \#1.}
	\label{fig:trispecOVM1}
\end{figure*}

\begin{figure*}[htbp]
	\begin{center}
	\includegraphics[scale=1]{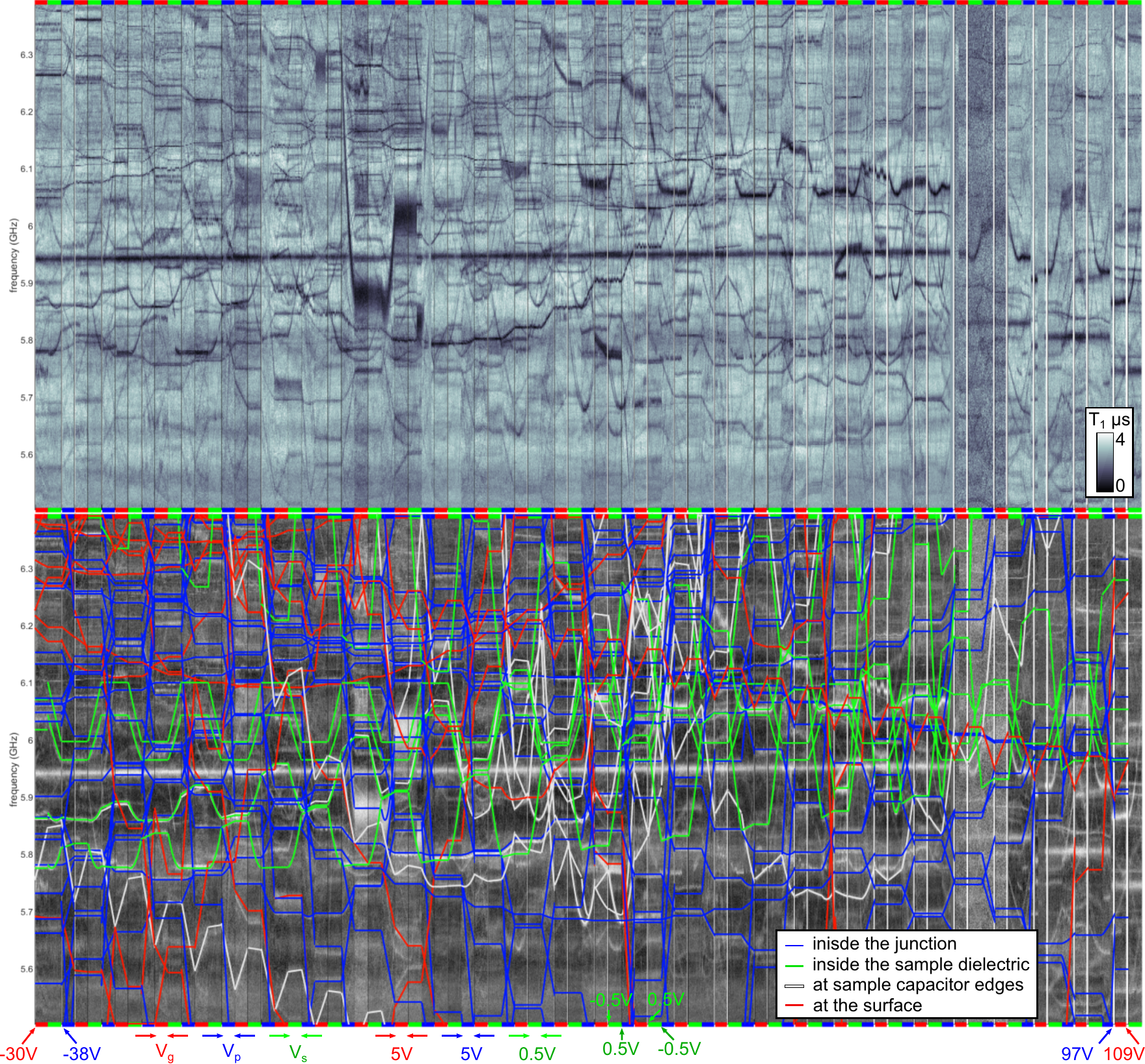}		
	\end{center}
	\caption{Top panel: Raw swap-spectroscopy data recorded on Qubit \#2 in a similar manner as Fig.~\ref{fig:trispecOVM1}. Bottom panel: same data with superimposed fits.}
	\label{fig:trispecOVM2}
\end{figure*}

Figure~\ref{fig:trispecOVM1} and Fig.~\ref{fig:trispecOVM2} show complete data sets recorded on nominally identically Qubits \#1 and \#2, respectively. 

\section*{Supplementary notes}
The here discussed work was part of the PhD studies of Alexander Bilmes at Karlsruhe Institute of Technology (KIT). Further details on the experimental setup, electric field simulations, and data acquisition can be found in his thesis~\cite{Bilmes19}.

\clearpage
\section*{Supplementary References}

\end{document}